\documentclass[11pt]{article}

\usepackage[margin=1in]{geometry}
\usepackage{amsmath}
\usepackage{amssymb}
\usepackage{graphicx}
\usepackage{booktabs}
\usepackage{float}
\usepackage{caption}
\usepackage{subcaption}
\usepackage{xcolor}
\usepackage[hidelinks]{hyperref}
\usepackage{tikz}
\usetikzlibrary{positioning,arrows.meta,shapes.geometric,fit,backgrounds}

\title{Stochastics of News Article Propagation: \\
       Joint Modeling of Counts and Reliability Composition \\
       in Event-Driven News Cascades}
\author{Harry Wang \and Sriyan Madugula \and Ruthesh Thavamani}
\date{MATH 460 --- Stochastic Processes}

\newcommand{\Npos}{N_{\mathrm{pos}}}
\newcommand{\Nneg}{N_{\mathrm{neg}}}
\newcommand{\Ntot}{N_{\mathrm{tot}}}
\newcommand{\pipos}{\pi_{\mathrm{pos}}}
\newcommand{\lampos}{\lambda_{\mathrm{pos}}}
\newcommand{\lamneg}{\lambda_{\mathrm{neg}}}

\begin{document}
\maketitle

\begin{abstract}
We study the news cascade triggered by a single high-impact event as a pair of
coupled stochastic processes: the daily article count $N(t)$ and the
reliable-vs-unreliable composition $\pipos(t)$. Using the 2017 Las Vegas
shooting as a primary case study (CC-News + NewsGuard, 91 days, 2{,}702
articles), we fit and compare four counting-process models for $N(t)$ and four
distributional models for $\pipos(t)$ on a common likelihood scale. For
counts, a hybrid Inhomogeneous-Poisson plus Hawkes model wins decisively
on AIC, BIC, and log-likelihood, as the IHP accounts for the initial exogenous shock,
while the Hawkes accounts for the subsequent self-excitation. A profile-likelihood
sweep of the Hawkes decay rate collapses the optimal Hawkes kernel to an
AR(1) one-step lag $\lambda(t)=\mu+n N(t-1)$. This Markovian structure on the
counts directly motivates the Part-B reliability models: we treat the
reliability label as a first-order Markov chain on a binary state space and
test homogeneous, regime-switching, and mean-field parameterizations against a
bivariate Hawkes baseline model using a shared conditional binomial likelihood.
The results show that the bivariate Hawkes model and the non-homogeneous Markov models 
are nearly indistinguishable on fit, as they exhaust the signal and start fitting to the noise. A
brief replication on the 2017 Hurricane Harvey data shows that the same methodology applies
but produces different results due to Hurricane Harvey's gradual endogenous build-up.
\end{abstract}

\section{Introduction}

A single high-impact event causes a cascade of news articles. Two observable
quantities of that cascade are simultaneously stochastic: \emph{how much} is
published on each day after the event ($N(t)$) and \emph{how reliable} that
coverage is ($\pipos(t)$, the fraction of articles published by reliable
outlets). Most existing work treats one of these in isolation. We argue that
both deserve a joint stochastic-process treatment on the same event timeline,
and that doing so is feasible with public data.

\paragraph{Research questions.}
\begin{itemize}
  \item \textbf{Part A.} What stochastic process best fits the daily article
        count $N(t)$ following an event?
  \item \textbf{Part B.} What process governs the reliable-vs-unreliable
        composition $\pipos(t)$ over the same window?
\end{itemize}

\paragraph{Running example.} The 2017 Las Vegas shooting is our primary case
study. The event has a singular, exogenous, well-defined origin (no prior
coverage ramp-up), a sharp day-0 spike (489 articles), and a long, decaying
tail (91 days). Reliable and unreliable outlets both engaged with the story so the
reliability composition is non-trivial. However, it should be noted that the there
are a majority of reliable outlets (63.5\%), which could result in some model biases. 
Hurricane Harvey 2017 serves as a secondary cross-event replication.

\paragraph{Prior work.} Three threads in the literature are relevant.

\emph{Point-process models of online attention.} Crane and Sornette
\cite{CraneSornette2008} used Hawkes-type response functions to separate
endogenous from exogenous bursts in YouTube views, which is conceptually
similar to our Part-A hybrid. Zhao et al.\ (SEISMIC, \cite{Zhao2015SEISMIC})
and Mishra, Rizoiu, and Xie \cite{MishraRizoiuXie2016} extended self-exciting
processes to tweet/retweet cascades. Daley and Vere-Jones
\cite{DaleyVereJones2003} provide the canonical reference for inhomogeneous
Poisson and Hawkes processes on point patterns.

\emph{Misinformation and reliability.} Vosoughi, Roy, and Aral
\cite{VosoughiRoyAral2018} showed descriptively that false news travels faster
and farther on Twitter than true news, but did not write down a generative
stochastic model. Allcott and Gentzkow \cite{AllcottGentzkow2017} measured the
economics of fake news during the 2016 U.S.\ election. Tambuscio et al.\
\cite{Tambuscio2015} used SIR-style compartmental models for hoax diffusion.
This is closest in spirit to our Markov-chain work, but in the context of continuous-time
epidemiology rather than discrete-time time-series.

\emph{Reliability scoring infrastructure.} NewsGuard and Media Bias / Fact
Check provide outlet-level reliability scores on a scale from 0 to 100. To convert these 
scores to a binary label, we threshold at the midpoint so that scores above 50 are considered
reliable and scores below 50 are considered unreliable. These scores are typically used as
features for downstream classification, rarely as the response variable in a
stochastic time-series model. We use NewsGuard explicitly as the response.

\paragraph{Gap and contributions.} We are not aware of prior work that
provides a joint, event-level, stochastic-process treatment of \emph{both} the
count and the reliability composition on the \emph{same} timeline, fit with
the \emph{same} statistical scoring. Thus, our paper attempts to contribute:
\begin{enumerate}
  \item A hybrid IHP\,+\,Hawkes model for the count process that decomposes a
        news cascade into an exogenous trigger and endogenous amplification
        operating at two distinct timescales.
  \item An empirical bridge from Part A to Part B: the Hawkes profile
        likelihood collapses to an AR(1) limit, motivating Markov-chain
        modeling of the reliability state. This narrative thread is the
        backbone of \S\ref{sec:bridge}.
  \item A shared conditional binomial likelihood that puts a multivariate
        Hawkes process and a family of Markov-chain models on a single AIC/BIC
        scale, despite having native likelihoods of different types. This allows
        us to directly compare how these models perform on the same task of predicting 
        the reliability composition.
  
\end{enumerate}

\section{Methods}
\label{sec:methods}

\subsection{Data and preprocessing}
\label{sec:data}

We use the news subset of the Common Crawl (CC-News), which provides URL,
publication date, and article body text. We left-join against the public
NewsGuard domain-level reliability dataset on the article's URL domain
(see \texttt{loader.py}), so that each article inherits the score of its
publishing outlet. The continuous NewsGuard score is thresholded at the
median to yield a binary reliability label $\{-1,+1\}$ (unreliable /
reliable). Event-relevant articles are selected by a keyword filter
(per-event TOML configuration); missing dates are zero-filled so every model
sees the same regular daily grid.

For the Las Vegas 2017 main analysis the window is 2017-10-02 to 2017-12-31
($T=91$ days, zero-filled), with $2{,}702$ articles total
($\Nneg=986$, $\Npos=1{,}716$; empirical split $0.365/0.635$), a peak of
$489$ articles on day 0, and $5$ days with zero articles (skipped in the
binomial likelihood).

\subsection{Part A: counting-process models for $N(t)$}
\label{sec:partA-models}

We fit three baseline counting-process models and a four-parameter
hybrid. All four models are scored on the same conditional Poisson
log-likelihood including the $\log N!$ normalization,
\begin{equation}
  \ell = \sum_{t=0}^{T-1}\Bigl[N(t)\log\lambda(t) - \lambda(t) - \log N(t)!\Bigr],
  \label{eq:poisson-LL}
\end{equation}
so that AIC and BIC are directly comparable.

\paragraph{Standard Poisson (SP, 1 parameter).}
$\lambda(t)=\lambda$ for all $t$; MLE in closed form,
$\hat\lambda=\overline{N}$. Included as a null baseline.

\paragraph{Inhomogeneous Poisson (IHP, 3 parameters).}
$\lambda(t)=A\,e^{-\beta t}+c$ with $A\!\geq\!0$ the initial amplitude above
baseline, $\beta\!>\!0$ the exponential decay rate, and $c\!\geq\!0$ the
long-run baseline. Each day's count is an independent Poisson draw whose rate
decays deterministically with absolute calendar time: this captures the
``news-cycle decay'' phenomenon. Optimized by Nelder--Mead with multi-start.

\paragraph{Discrete-time Hawkes (2 parameters in the AR-1 limit, 3 in general).}
We use the recursive form
\begin{equation}
  \lambda(t)=\mu+\alpha\,R(t),\qquad R(t)=e^{-\beta}\bigl[R(t-1)+N(t-1)\bigr],
  \quad R(0)=0,
  \label{eq:hawkes}
\end{equation}
with the stationary reparameterization
\begin{equation}
  n=\frac{\alpha}{e^{\beta}-1}\in(0,1),
\end{equation}
where $n$ is the \emph{branching ratio} (the expected number of secondary
articles triggered per article). $\lambda(t)$ is conditional on the past
counts $N(0),\dots,N(t-1)$, so the log-likelihood
\eqref{eq:poisson-LL} must be evaluated recursively rather than with a
close-form MLE on independent Poisson counts.

\paragraph{Hybrid IHP\,+\,Hawkes (AR-1, 4 parameters).}
The hybrid combines exogenous decay with endogenous self-excitation:
\begin{equation}
  \lambda(0)=A+c,\qquad
  \lambda(t)=A\,e^{-\beta_0 t}+c+n\,N(t-1),\quad t\ge 1.
\end{equation}
The $A\,e^{-\beta_0 t}+c$ term captures exogenous news-cycle decay, while the
$n\,N(t-1)$ term captures endogenous next-day amplification. Hybrid-AR1
strictly nests both pure IHP (set $n=0$) and pure Hawkes-AR1 (drop the IHP
term and rename $c\!\to\!\mu$), enabling likelihood-ratio tests in addition to
AIC/BIC.

\paragraph{Optimization.} All models are fit with Nelder--Mead from a
multi-start grid of initializations to dodge local minima. Stationarity of
the Hawkes process is checked at the fitted value; for the multivariate
Hawkes of \S\ref{sec:partB-models} stationarity is enforced as a
spectral-radius constraint on the branching matrix.

\subsection{The AR(1)\,$\to$\,Markov bridge between Part A and Part B}
\label{sec:bridge}

An important result of the Part-A Hawkes model that profoundly shapes Part B is the
$\beta$-profile sweep. Fixing $\beta$ on
a log-grid from $0.05$ to $32$ and re-optimizing $(\mu,n)$, the profile
log-likelihood is monotonically increasing in $\beta$ and plateaus for
$\beta\geq 10$ (Table~\ref{tab:beta-profile}). In the limit
$\beta\to\infty$ the exponential kernel $e^{-\beta\Delta}$ collapses to a
delta at lag $\Delta=1$, and the Hawkes intensity reduces to
\begin{equation}
  \lambda(t)\approx \mu + n\,N(t-1).
  \label{eq:ar1}
\end{equation}
This results in a discrete AR(1) model for the count
process: the next-day depends only on yesterday's count, which is
precisely the structure of a first-order Markov chain.

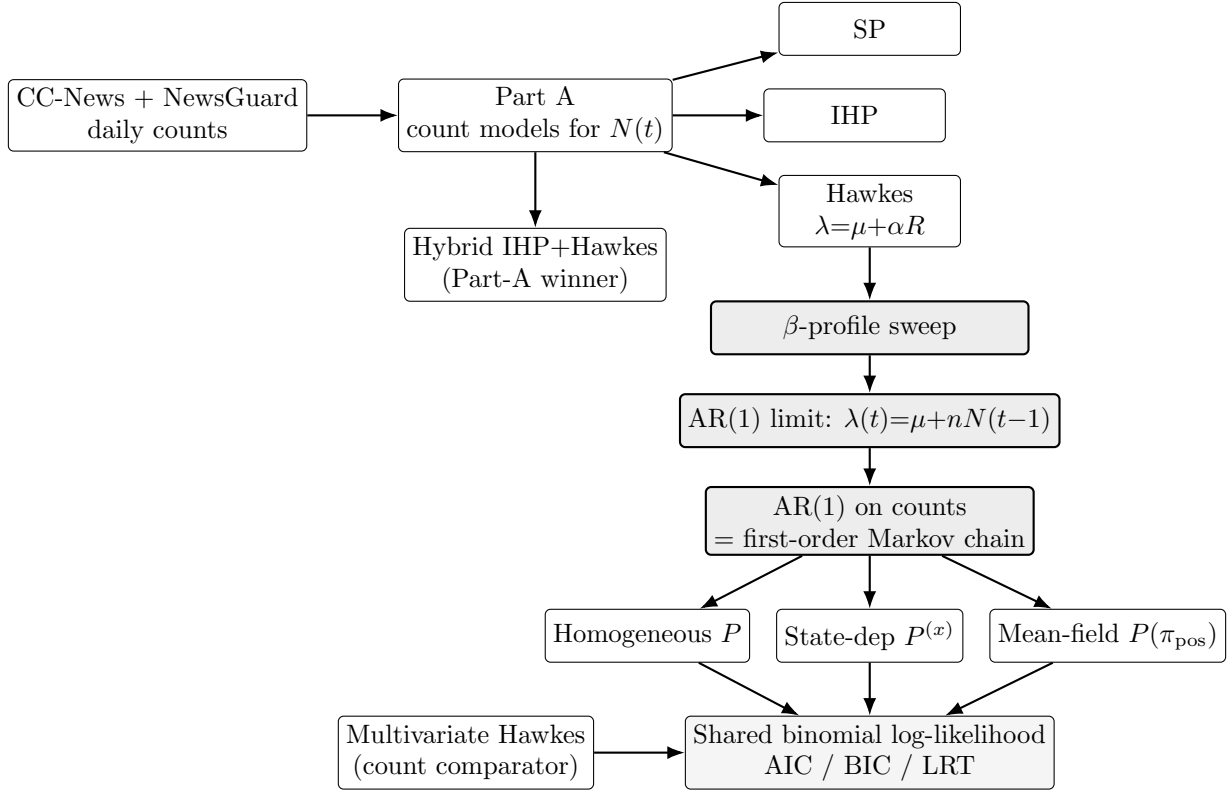
\begin{figure}[H]
\centering
\begin{tikzpicture}[
    node distance=6mm and 12mm,
    every node/.style={align=center, font=\small},
    box/.style={draw, rectangle, rounded corners=2pt, inner sep=3pt,
                minimum height=7mm, minimum width=24mm},
    pivot/.style={draw, rectangle, rounded corners=2pt, inner sep=3pt,
                  minimum height=7mm, minimum width=42mm,
                  thick, fill=black!7},
    arrow/.style={-Latex, thick}
]
  \node[box] (data) {CC-News + NewsGuard \\ daily counts};
  \node[box, right=of data] (partA) {Part A \\ count models for $N(t)$};
  \node[box, above right=3mm and 14mm of partA] (sp)     {SP};
  \node[box, right=of partA]                    (ihp)    {IHP};
  \node[box, below right=3mm and 14mm of partA] (hawkes) {Hawkes \\ $\lambda{=}\mu{+}\alpha R$};
  \node[box, below=10mm of partA] (hybrid) {Hybrid IHP+Hawkes \\ (Part-A winner)};
  \node[pivot, below=7mm of hawkes] (betaSweep) {$\beta$-profile sweep};
  \node[pivot, below=5mm of betaSweep] (ar1)
        {AR(1) limit: $\lambda(t){=}\mu{+}n N(t{-}1)$};
  \node[pivot, below=5mm of ar1] (bridge)
        {AR(1) on counts \\ = first-order Markov chain};
  \node[box, below left=7mm and -6mm of bridge] (homog)     {Homogeneous $P$};
  \node[box, below=7mm of bridge]                (stateDep)  {State-dep $P^{(x)}$};
  \node[box, below right=7mm and -6mm of bridge](meanField) {Mean-field $P(\pipos)$};
  \node[box, below=7mm of stateDep, fill=black!4] (binomLL)
        {Shared binomial log-likelihood \\ AIC / BIC / LRT};
  \node[box, left=of binomLL] (mvHawkes) {Multivariate Hawkes \\ (count comparator)};

  \draw[arrow] (data) -- (partA);
  \draw[arrow] (partA) -- (sp);
  \draw[arrow] (partA) -- (ihp);
  \draw[arrow] (partA) -- (hawkes);
  \draw[arrow] (partA) -- (hybrid);
  \draw[arrow] (hawkes) -- (betaSweep);
  \draw[arrow] (betaSweep) -- (ar1);
  \draw[arrow] (ar1) -- (bridge);
  \draw[arrow] (bridge) -- (homog);
  \draw[arrow] (bridge) -- (stateDep);
  \draw[arrow] (bridge) -- (meanField);
  \draw[arrow] (mvHawkes) -- (binomLL);
  \draw[arrow] (homog)     -- (binomLL);
  \draw[arrow] (stateDep)  -- (binomLL);
  \draw[arrow] (meanField) -- (binomLL);
\end{tikzpicture}
\caption{The narrative thread connecting Part A and Part B. The Hawkes
$\beta$-profile sweep on daily-resolution counts plateaus in the AR(1)
limit, and an AR(1) on a discrete state is precisely a first-order Markov
chain. This motivates a Markov-chain family for the binary reliability
state in Part B, scored against the multivariate-Hawkes model on a
common conditional binomial likelihood.}
\label{fig:ar1-bridge}
\end{figure}

This empirical AR(1) collapse motivates the Part-B model family.
For the binary reliability state, the natural generalization of
\eqref{eq:ar1} is a first-order Markov chain on the state space
$\{\mathrm{neg},\mathrm{pos}\}$, parameterized by a $2{\times}2$ transition
matrix $P$ that may be \emph{homogeneous} (a single $P$), \emph{regime-switching}
(two $P^{(\cdot)}$ matrices indexed by the previous day's majority state), or
\emph{mean-field} (the entries of $P_t$ depend on yesterday's
empirical proportion $\pipos(t-1)$). This is summarized in
Figure~\ref{fig:ar1-bridge}.

\subsection{Part B: distributional models for the reliability split}
\label{sec:partB-models}

Part B asks: given the daily total $\Ntot(t)=\Nneg(t)+\Npos(t)$, what process
explains the daily \emph{reliable fraction} $\pipos(t)=\Npos(t)/\Ntot(t)$?

\paragraph{Shared conditional binomial likelihood.}
Hawkes natively scores the pair of counts $(\Nneg(t),\Npos(t))$ under a
multivariate Poisson log-likelihood, and a Markov chain natively scores a
one-step transition under a multinomial / binomial on the proportion. The
two log-likelihoods live on different scales, so direct comparison is
meaningless. To put all the metrics on one scale we use the
Poisson--binomial superposition identity: if
$X\sim\mathrm{Pois}(\lambda_1)$ and $Y\sim\mathrm{Pois}(\lambda_2)$
independently, then
$X\mid X+Y=n \sim \mathrm{Binomial}(n, \lambda_1/(\lambda_1+\lambda_2))$.
Applied to our setup,
\begin{equation}
  \Npos(t)\,\bigl|\,\Ntot(t)\;\sim\;\mathrm{Binomial}\bigl(\Ntot(t),\,q_t\bigr),
  \label{eq:cond-binomial}
\end{equation}
with model-specific $q_t$. The per-day log-likelihood is
\begin{equation}
  \ell_t = \log\binom{\Ntot(t)}{\Npos(t)}
         + \Npos(t)\log q_t + \Nneg(t)\log(1-q_t),
  \label{eq:binom-LL}
\end{equation}
summed over the $86$ days with $\Ntot(t)>0$. The binomial coefficient is a
constant in $q_t$ and so cancels in every $\Delta\mathrm{AIC}$ and LRT.

\paragraph{Model 1: Multivariate Hawkes (AR-1 limit, 6 parameters).}
\begin{equation}
  \lambda_k(t) = \mu_k + \sum_{j\in\{\mathrm{neg},\mathrm{pos}\}} n_{kj}\,N_j(t-1),
  \quad k\in\{\mathrm{neg},\mathrm{pos}\}.
  \label{eq:mvhawkes}
\end{equation}
Each category gets its own baseline $\mu_k$ and its own row of the
$2{\times}2$ branching matrix $n=(n_{kj})$, so cross-excitation is
first-class. Stationarity ($\rho(n)<1$) is enforced as a constraint during
optimization. The implied split is
$q_t=\lampos(t)/(\lampos(t)+\lamneg(t))$, the direct consequence of the
superposition identity \eqref{eq:cond-binomial}.

\paragraph{Model 2a: Homogeneous Markov (null, 2 parameters).}
A single $2{\times}2$ transition matrix $P$ with rows summing to one
(only two free parameters, one per row). The predicted split is
$q_t=(\pi_{t-1}\,P)_{\mathrm{pos}}$.

\paragraph{Model 2b: State-dependent Markov (4 parameters).}
Two transition matrices $P^{(\mathrm{neg})}$ and $P^{(\mathrm{pos})}$,
indexed by yesterday's majority class
$x_{t-1}=\arg\max_k\pi_k(t-1)\in\{\mathrm{neg},\mathrm{pos}\}$:
\begin{equation}
  q_t = \bigl(\pi_{t-1}\, P^{(x_{t-1})}\bigr)_{\mathrm{pos}}.
\end{equation}
This lets the rare neg-majority shock regime have qualitatively different
dynamics from the typical pos-majority equilibrium regime.

\paragraph{Model 2c: Mean-field Markov (4 parameters).}
The transition probabilities depend smoothly on yesterday's empirical
proportion through logistic links,
\begin{align}
  P_{\mathrm{neg}\to\mathrm{pos}}(t) &= \sigma\bigl(a_0 + a_1\,\pipos(t-1)\bigr),
   \\
  P_{\mathrm{pos}\to\mathrm{neg}}(t) &= \sigma\bigl(b_0 + b_1\,\pipos(t-1)\bigr),
\end{align}
with the other two entries of $P_t$ fixed by the row-sum constraint.

\paragraph{Optimization and validation.}
We again use a multi-start Nelder--Mead procedure for parameter estimation.
To validate the approach, we conducted a synthetic parameter-recovery experiment
for the multivariate Hawkes model ($T=500$), with
$\mu_{\mathrm{neg}}=3.0$, $\mu_{\mathrm{pos}}=5.0$, and a $2\times2$
branching matrix. Across runs, the multi-start procedure recovered all six
parameters with a maximum relative error of $20.3\%$, which we considered
sufficient for downstream inference. By contrast, a single-start Nelder--Mead
fit performed substantially worse on the same data (maximum relative error
$148\%$), highlighting the importance of multi-start initialization.

\subsection{Reproducibility checklist}
\label{sec:repro}

\begin{sloppypar}
The full analysis is reproducible from the provided repository:
\texttt{loader.py} ingests CC-News and joins NewsGuard;
\mbox{\texttt{partAresults/fit\_models.py}} fits SP, IHP, and Hawkes;
\mbox{\texttt{partAresults/fit\_hybrid.py}} fits the hybrid;
\mbox{\texttt{partBresults/fit\_models\_partB.py}} fits all four Part-B
candidates and produces all figures referenced below. Per-event
configurations live in \texttt{events/*.toml}; the published code
intentionally exposes the multi-start grid so that the synthetic-recovery
diagnostic is reproducible on demand.
\end{sloppypar}

\section{Results}
\label{sec:results}

\subsection{Part A on Las Vegas 2017: counts}
\label{sec:results-partA-LV}

\begin{figure}[H]
\centering
\includegraphics[width=0.85\linewidth]{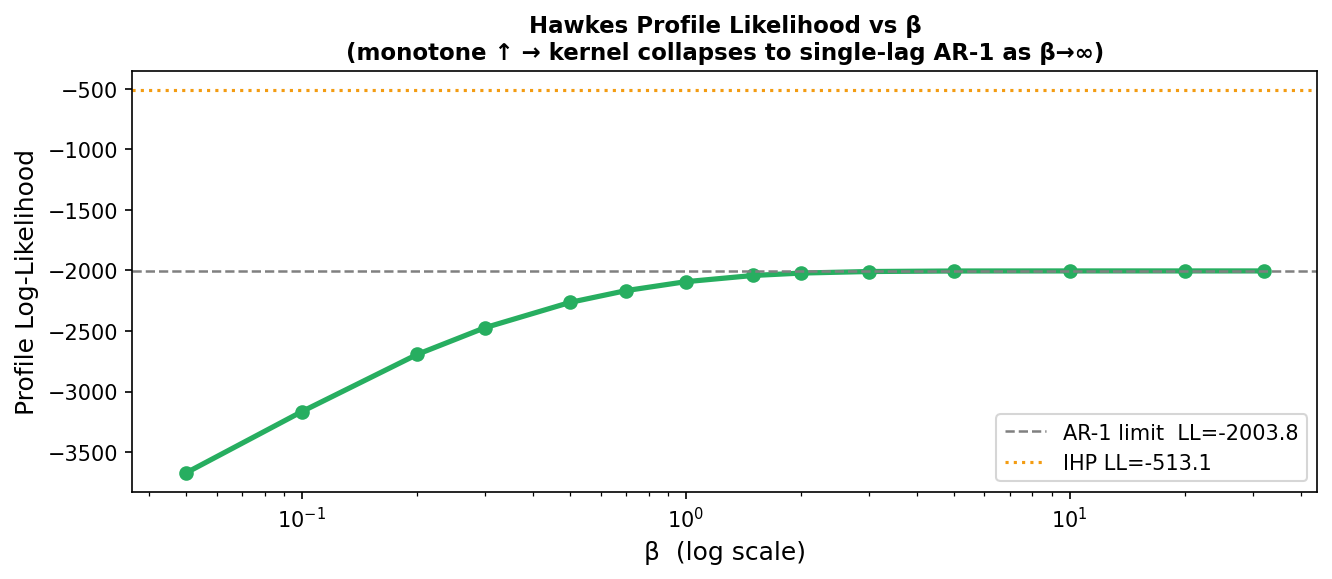}
\caption{Hawkes profile log-likelihood vs.\ kernel decay $\beta$ on Las
Vegas 2017. Monotone in $\beta$, plateauing for $\beta\!\geq\!10$. The
plateau is the AR(1) limit of \eqref{eq:ar1} with
$\hat\mu\approx 11.50$, $\hat n\approx 0.614$.}
\label{fig:partA-beta-profile}
\end{figure}

\paragraph{Profile sweep and AR(1) collapse.}
The Hawkes profile-likelihood sweep over $\beta$ (Table~\ref{tab:beta-profile})
is monotonically increasing in $\beta$ and stable to four decimal places for
$\beta\geq 10$. The plateau value
$(\hat\mu,\hat n)=(11.50, 0.6137)$ defines the AR(1) Hawkes we report as
Model C in Table~\ref{tab:partA-models}. Figure~\ref{fig:partA-beta-profile}
visualizes the sweep.

\begin{table}[H]
\centering
\caption{Profile log-likelihood of the discrete-time Hawkes model on Las
Vegas 2017 over a log-grid in the kernel decay $\beta$.
Re-optimization of $(\mu,n)$ at each fixed $\beta$.}
\label{tab:beta-profile}
\begin{tabular}{rrrr}
\toprule
$\beta$ & $\hat\mu$ & $\hat n$ & Log-likelihood \\
\midrule
$0.05$  & $17.66$ & $0.424$ & $-3670.5$ \\
$0.10$  & $12.29$ & $0.595$ & $-3165.5$ \\
$0.20$  & $11.15$ & $0.630$ & $-2692.5$ \\
$0.30$  & $11.02$ & $0.633$ & $-2472.4$ \\
$0.50$  & $11.07$ & $0.630$ & $-2263.5$ \\
$1.00$  & $11.25$ & $0.623$ & $-2092.6$ \\
$2.00$  & $11.40$ & $0.617$ & $-2021.4$ \\
$5.00$  & $11.50$ & $0.614$ & $-2004.4$ \\
$10.00$ & $11.50$ & $0.614$ & $-2003.8$ \\
$20.00$ & $11.50$ & $0.614$ & $-2003.8$ \\
$32.00$ & $11.50$ & $0.614$ & $-2003.8$ \\
\bottomrule
\end{tabular}
\end{table}

\paragraph{Model comparison.} Table~\ref{tab:partA-models} compares all
four models on AIC, BIC, and log-likelihood.

\begin{table}[H]
\centering
\caption{Part A: model comparison on the Las Vegas 2017 cascade
($T=91$ days).}
\label{tab:partA-models}
\begin{tabular}{lrrrrr}
\toprule
Model & \# Params & Log-Lik. & AIC & BIC \\
\midrule
Standard Poisson         & $1$ & $-3835.11$ & $7672.22$ & $7674.73$ \\
Inhomogeneous Poisson    & $3$ & $-513.15$  & $1032.29$ & $1039.82$   \\
Hawkes (AR-1 limit)      & $2$ & $-2003.81$ & $4011.62$ & $4016.65$ \\
\textbf{Hybrid IHP+Hawkes (AR-1)} & $\mathbf{4}$ & $\mathbf{-388.79}$  & $\mathbf{785.59}$  & $\mathbf{795.63}$   \\
\bottomrule
\end{tabular}
\end{table}

\paragraph{Likelihood-ratio tests.} Both pure models are overwhelmingly
preferred to Standard Poisson; the hybrid additions are also significant.

\begin{table}[H]
\centering
\caption{Likelihood-ratio tests on Las Vegas 2017. Each comparison nests
the smaller model strictly.}
\label{tab:partA-LRT}
\begin{tabular}{lrrl}
\toprule
Comparison & LR statistic & df & $p$-value \\
\midrule
SP vs.\ IHP                  & $6{,}643.9$ & $2$ & $<10^{-300}$ \\
SP vs.\ Hawkes               & $3{,}662.6$ & $1$ & $<10^{-300}$ \\
IHP vs.\ Hybrid-AR1          & $248.7$     & $1$ & $\approx 0$  \\
\bottomrule
\end{tabular}
\end{table}

\paragraph{Why the hybrid wins: the day-0 decomposition.}
Pure Hawkes loses to IHP (and to the hybrids) almost entirely because of
day~0. With $R(0)=0$, pure Hawkes predicts
$\lambda(0)=\mu=11.5$ against the observation $N(0)=489$, contributing
$-1{,}360.13$ to the log-likelihood; $\approx 91\%$ of the entire
Hawkes--IHP gap of $1490.7$. The IHP and hybrid models can absorb the
exogenous shock through the $A\,e^{-\beta_0\cdot 0}=A$ term:
\begin{center}
\begin{tabular}{lcr}
\toprule
Model & $\lambda(0)$ & Day-0 log-lik \\
\midrule
IHP                    & $450.6$ & $-5.60$    \\
Hawkes (AR-1)          & $11.5$  & $-1360.13$ \\
Hybrid (AR-1)          & $489.0$ & $-4.02$    \\
\bottomrule
\end{tabular}
\end{center}
Hybrid-AR1 lands $\lambda(0)$ \emph{exactly} on the observed $489$ articles
(the optimizer drives $A+c$ onto the data), then hands propagation off to a
self-excitation term with branching ratio $\hat n=0.7248$.

\begin{figure}[H]
\centering
\includegraphics[width=0.95\linewidth]{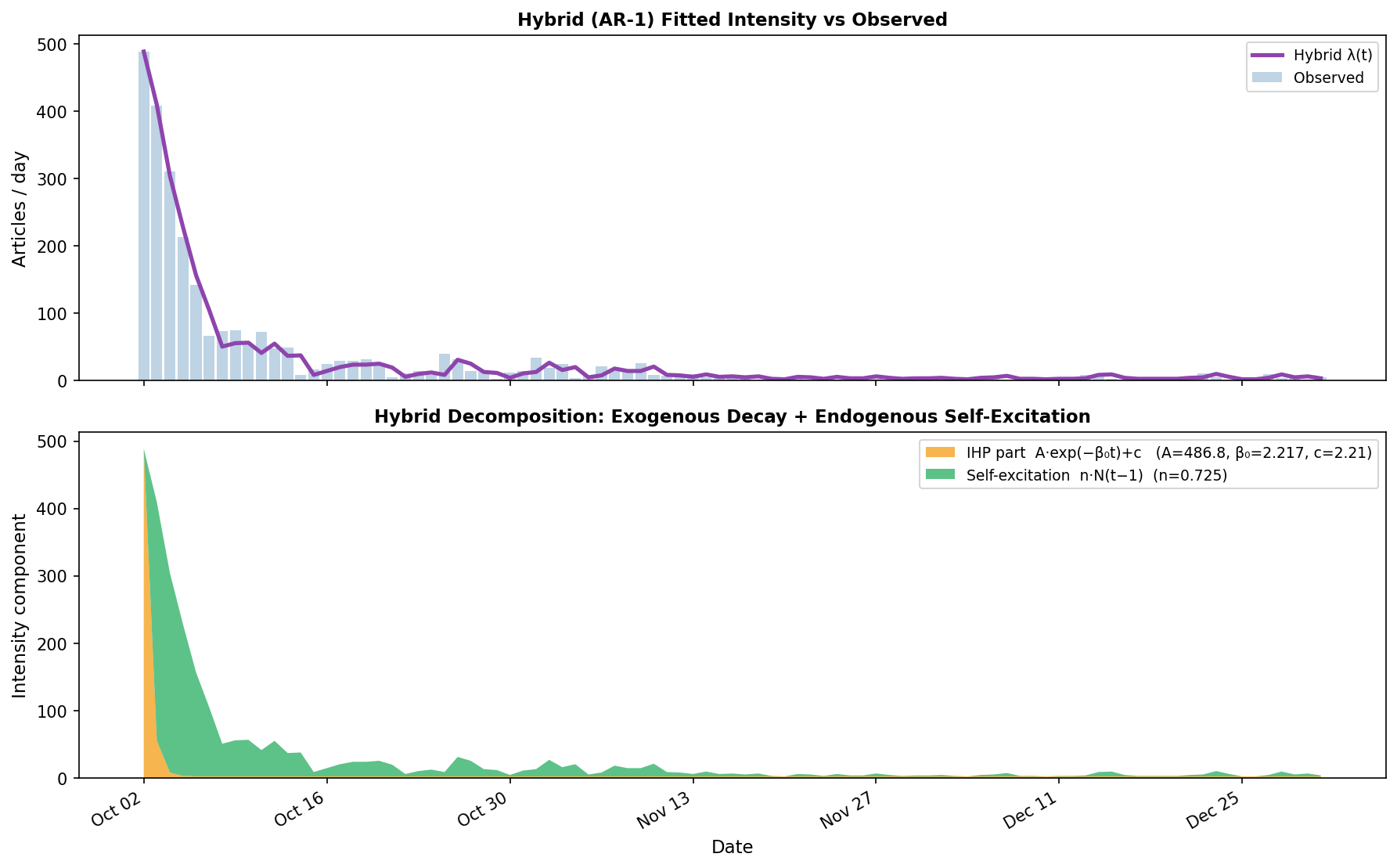}
\caption{Hybrid-AR1 decomposition on Las Vegas 2017. The IHP term collapses
to a near-delta spike that absorbs the day-0 burst (initial amplitude
$A\approx 487$, $\beta_0=2.22 \Rightarrow$ half-life $0.31$ days), and the
self-excitation term ($n=0.7248$) carries the subsequent endogenous
dynamics.}
\label{fig:partA-hybrid-decomp}
\end{figure}

\begin{figure}[H]
\centering
\includegraphics[width=0.95\linewidth]{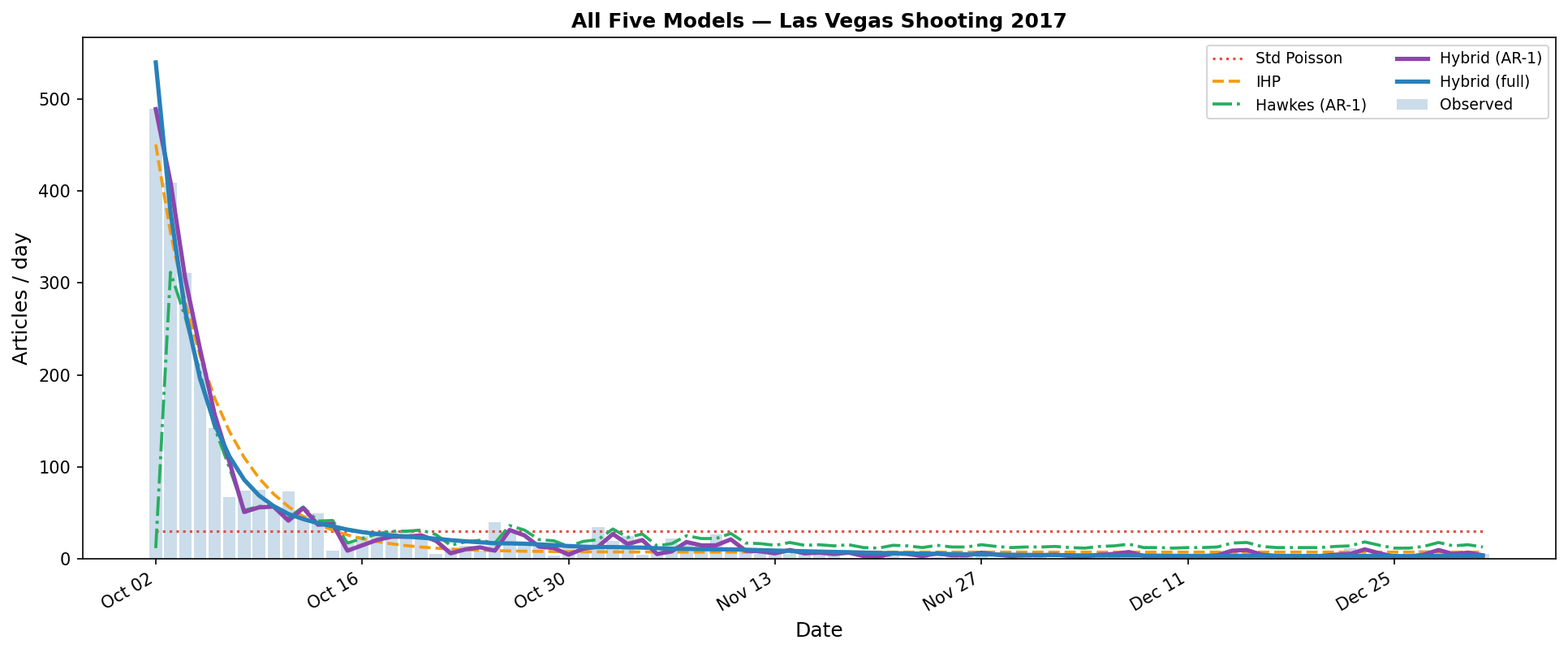}
\caption{All four Part-A models overlaid on the observed daily article
counts (Las Vegas 2017).}
\label{fig:partA-overlay}
\end{figure}

\paragraph{Hybrid mechanism decomposition.}
The AR-1 hybrid separates the cascade into an exogenous IHP component and an
endogenous self-excitation component. The fitted IHP half-life is
$\ln 2 / 2.22 \approx 0.31$ days for the sharp initial drop, while the AR-1
branching ratio ($\hat n=0.7248$) controls the persistence of the endogenous
tail.
Figure~\ref{fig:partA-hybrid-decomp} shows the decomposition of the hybrid model,
and Figure~\ref{fig:partA-overlay} overlays all four models against
the observed counts.

\subsection{Part B on Las Vegas 2017: reliability split}
\label{sec:results-partB-LV}

\paragraph{Model comparison.}
On the shared conditional binomial likelihood
\eqref{eq:binom-LL} over the $86$ non-empty days, the four candidate
models compare as in Table~\ref{tab:partB-models}.

\begin{table}[H]
\centering
\caption{Part B: model comparison on Las Vegas 2017. Common conditional
binomial log-likelihood.}
\label{tab:partB-models}
\begin{tabular}{lrrrr}
\toprule
Model & \# Params & Log-Lik. & AIC & BIC \\
\midrule
Multivariate Hawkes        & $6$ & $-156.93$ & $325.87$ & $340.60$ \\
Homogeneous Markov         & $2$ & $-162.75$ & $329.50$ & $334.41$ \\
\textbf{State-dependent Markov} & $\mathbf{4}$ & $\mathbf{-157.33}$ & $\mathbf{322.65}$ & $\mathbf{332.47}$ \\
Mean-field Markov          & $4$ & $-157.56$ & $323.13$ & $332.94$ \\
\bottomrule
\end{tabular}
\end{table}

The state-dependent Markov is the AIC winner. Mean-field Markov is a
statistically indistinguishable second ($\Delta\mathrm{AIC}=0.48$).
Multivariate Hawkes beats the homogeneous baseline on raw log-likelihood
(by $5.82$) but pays $+8$ in AIC for its $4$ extra parameters and so loses
on AIC.

\paragraph{Three near-tied non-null models.}
The headline of Table~\ref{tab:partB-models} is not a clean ranking but
rather two groupings. On raw log-likelihood the three non-trivial models
(Hawkes, state-dependent Markov, mean-field Markov) span only
$\approx 0.6$ nats ($-156.93$ to $-157.56$); pairwise they are essentially
indistinguishable on \emph{fit quality}. All three are $\approx\!5\text{--}6$
nats better than the homogeneous Markov null by a uniformly large gap that
the LRTs in Table~\ref{tab:partB-LRT} formalize for the two nested Markov
variants. AIC then breaks the within-group tie not by improved fit but by
\emph{parameter accounting}: state-dependent and mean-field Markov each
carry $4$ parameters, while multivariate Hawkes carries $6$, so Hawkes pays
a $+4$ AIC penalty for an $\approx 0.4$-nat LL improvement and ends up
behind. Thus, the three non-null models tie on fit; AIC selects the most
parameter-efficient parameterization among them.

\begin{table}[H]
\centering
\caption{Likelihood-ratio tests vs.\ the homogeneous Markov null on
Las Vegas 2017. Hawkes does not nest the Markov family, so AIC/BIC are
used for that comparison.}
\label{tab:partB-LRT}
\begin{tabular}{lrrl}
\toprule
Comparison & LR statistic & df & $p$-value \\
\midrule
State-dep Markov vs.\ Homogeneous  & $10.85$ & $2$ & $4.4\times 10^{-3}$ \\
Mean-field Markov vs.\ Homogeneous & $10.37$ & $2$ & $5.6\times 10^{-3}$ \\
\bottomrule
\end{tabular}
\end{table}

\begin{figure}[H]
\centering
\includegraphics[width=0.45\linewidth]{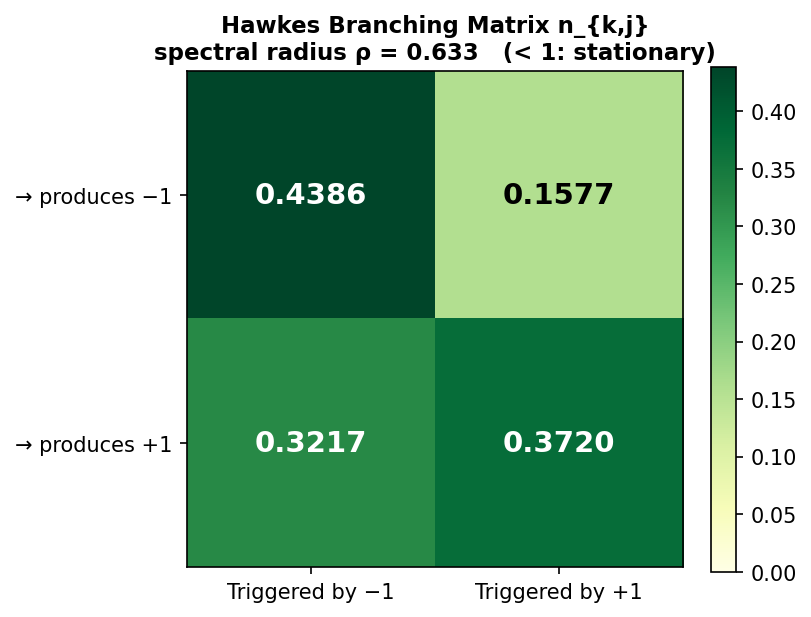}
\caption{Multivariate Hawkes branching matrix on Las Vegas 2017
(rows = triggered, columns = triggering). Diagonally dominant but not
overwhelmingly so; off-diagonals reveal genuine cross-excitation between
the reliable and unreliable streams.}
\label{fig:partB-branching}
\end{figure}

\paragraph{Multivariate Hawkes branching matrix.}
The fitted branching matrix on Las Vegas 2017
(rows = triggered category, columns = triggering category) is
\begin{equation}
  n =
  \begin{pmatrix}
    n_{\mathrm{neg},\mathrm{neg}} & n_{\mathrm{neg},\mathrm{pos}} \\
    n_{\mathrm{pos},\mathrm{neg}} & n_{\mathrm{pos},\mathrm{pos}}
  \end{pmatrix}
  =
  \begin{pmatrix}
    0.4386 & 0.1577 \\
    0.3217 & 0.3720
  \end{pmatrix},
  \qquad
  \mu = (3.12,\ 8.38),
\end{equation}
with spectral radius $\rho(n)=0.633$ (sub-critical).
The off-diagonals are non-negligible: an unreliable article today triggers
about $n_{\mathrm{pos},\mathrm{neg}}=0.32$ reliable articles tomorrow on
average (likely fact-checks, debunks, follow-ups), while a reliable article
triggers about $n_{\mathrm{neg},\mathrm{pos}}=0.16$ unreliable articles
tomorrow, with half as much in the reverse direction.
Figure~\ref{fig:partB-branching} visualizes the branching matrix as a
heatmap.

\begin{figure}[H]
\centering
\includegraphics[width=0.85\linewidth]{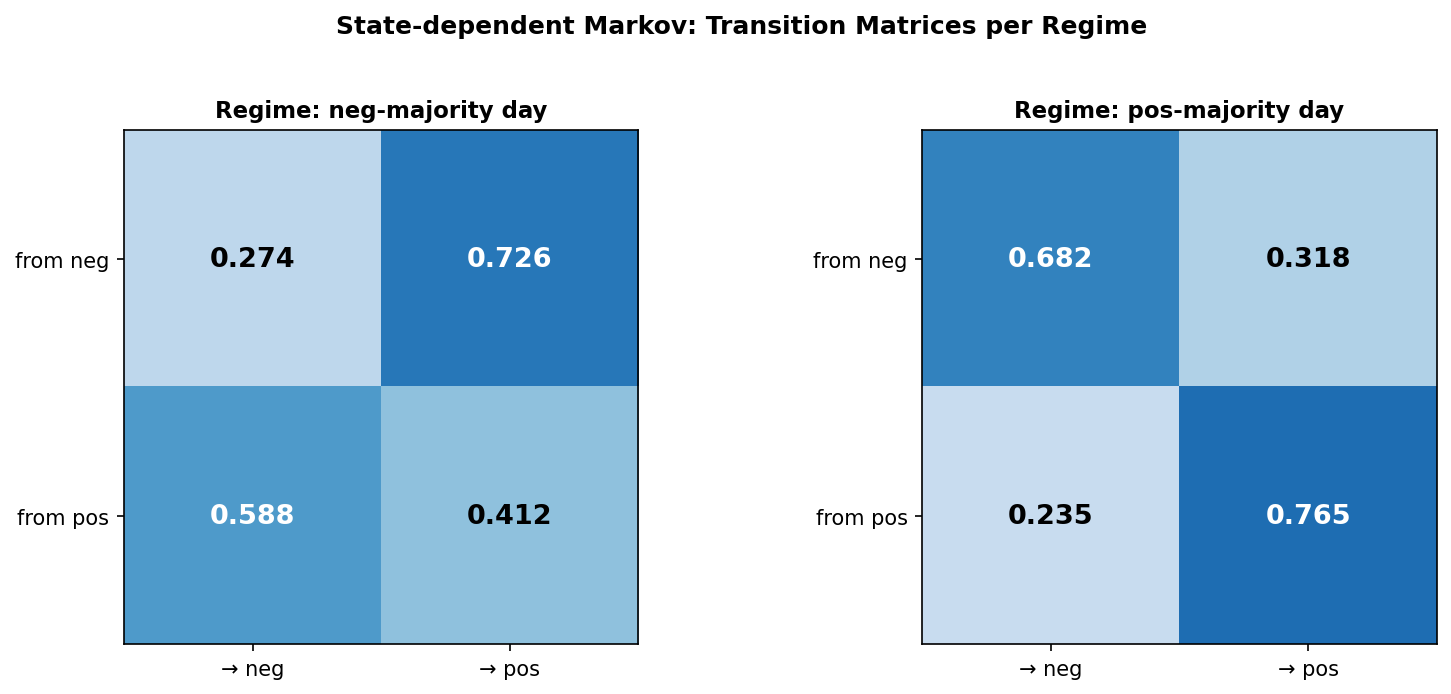}
\caption{State-dependent Markov on Las Vegas 2017: the two regime-conditional
$2{\times}2$ transition matrices. The pos-majority matrix is sticky on both
diagonals; the neg-majority matrix shows aggressive mean reversion toward
the reliable state.}
\label{fig:partB-regime-mats}
\end{figure}

\paragraph{Regime-switching Markov.}
The two regime-conditional transition matrices are:
\begin{align}
  P^{(\mathrm{neg})}
  &=
  \begin{pmatrix}
    0.2745 & 0.7255 \\
    0.5879 & 0.4121
  \end{pmatrix},
  \qquad
  P^{(\mathrm{pos})}
  =
  \begin{pmatrix}
    0.6818 & 0.3182 \\
    0.2354 & 0.7646
  \end{pmatrix}.
\end{align}
Rows are the source state (neg / pos), columns the destination. The two
regimes have qualitatively different dynamics:
\emph{pos-majority} is sticky on both diagonals
($P^{(\mathrm{pos})}_{\mathrm{neg}\to\mathrm{neg}}=0.68$,
$P^{(\mathrm{pos})}_{\mathrm{pos}\to\mathrm{pos}}=0.76$), while the rare
\emph{neg-majority} regime aggressively pushes back toward pos
($P^{(\mathrm{neg})}_{\mathrm{neg}\to\mathrm{pos}}=0.7255$,
$P^{(\mathrm{neg})}_{\mathrm{pos}\to\mathrm{neg}}=0.5879$). The system has a
strong restoring force toward its empirical equilibrium near
$\pipos\approx 0.63$. See Figure~\ref{fig:partB-regime-mats}.

\begin{figure}[H]
\centering
\includegraphics[width=0.95\linewidth]{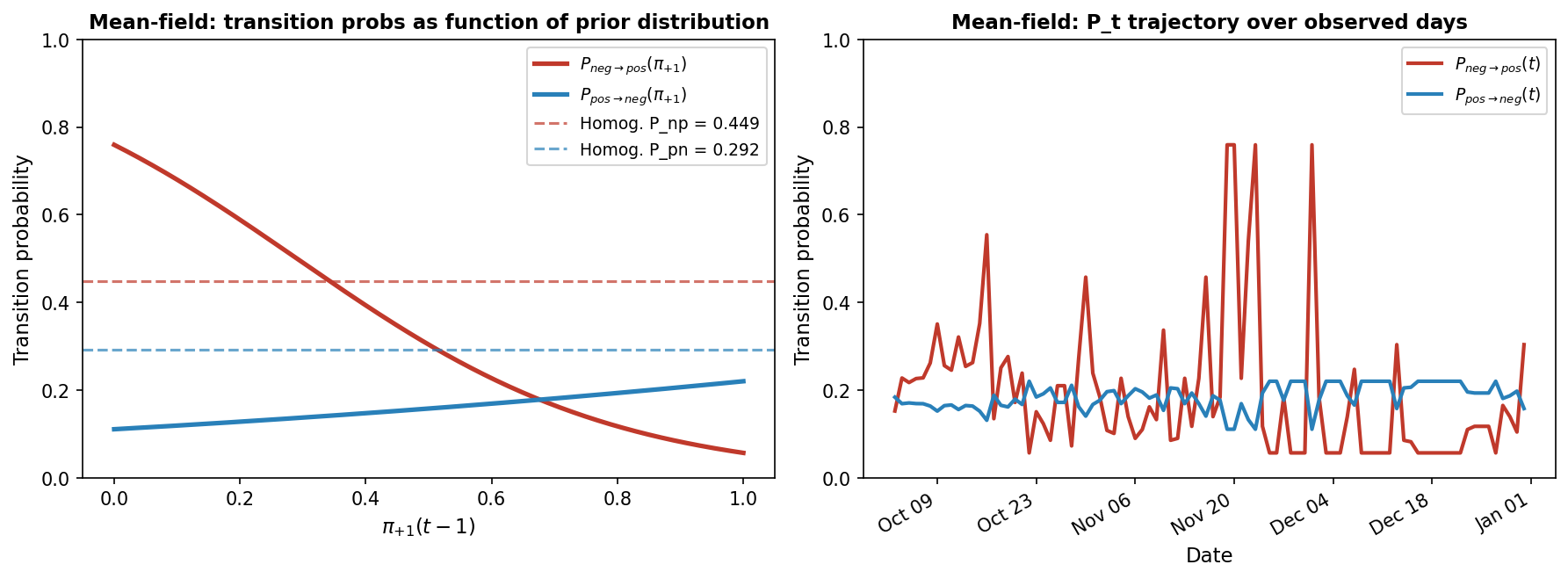}
\caption{Mean-field Markov on Las Vegas 2017: fitted logistic links
$P_{\mathrm{neg}\to\mathrm{pos}}(\pipos)$ and
$P_{\mathrm{pos}\to\mathrm{neg}}(\pipos)$, plus the resulting time
trajectory of the implied transition matrix entries.}
\label{fig:partB-meanfield}
\end{figure}

\paragraph{Mean-field Markov.}
Fitted parameters
$(a_0,a_1,b_0,b_1)=(+1.1496,-3.9587,-2.0828,+0.8184)$. At
$\pipos(t-1)=0$ the link gives
$P_{\mathrm{neg}\to\mathrm{pos}}=\sigma(1.15)=0.76$ (strong flip back to pos
when pos is absent); at $\pipos(t-1)=1$ it gives
$P_{\mathrm{neg}\to\mathrm{pos}}=\sigma(-2.81)=0.06$ (almost no flips when
pos already dominates). The decreasing logistic shape of
$P_{\mathrm{neg}\to\mathrm{pos}}$ as a function of yesterday's $\pipos$ is
the smooth analogue of the regime-switching mean reversion observed in 2b.
Figure~\ref{fig:partB-meanfield} plots the fitted curves and trajectory.

\begin{figure}[H]
\centering
\includegraphics[width=0.95\linewidth]{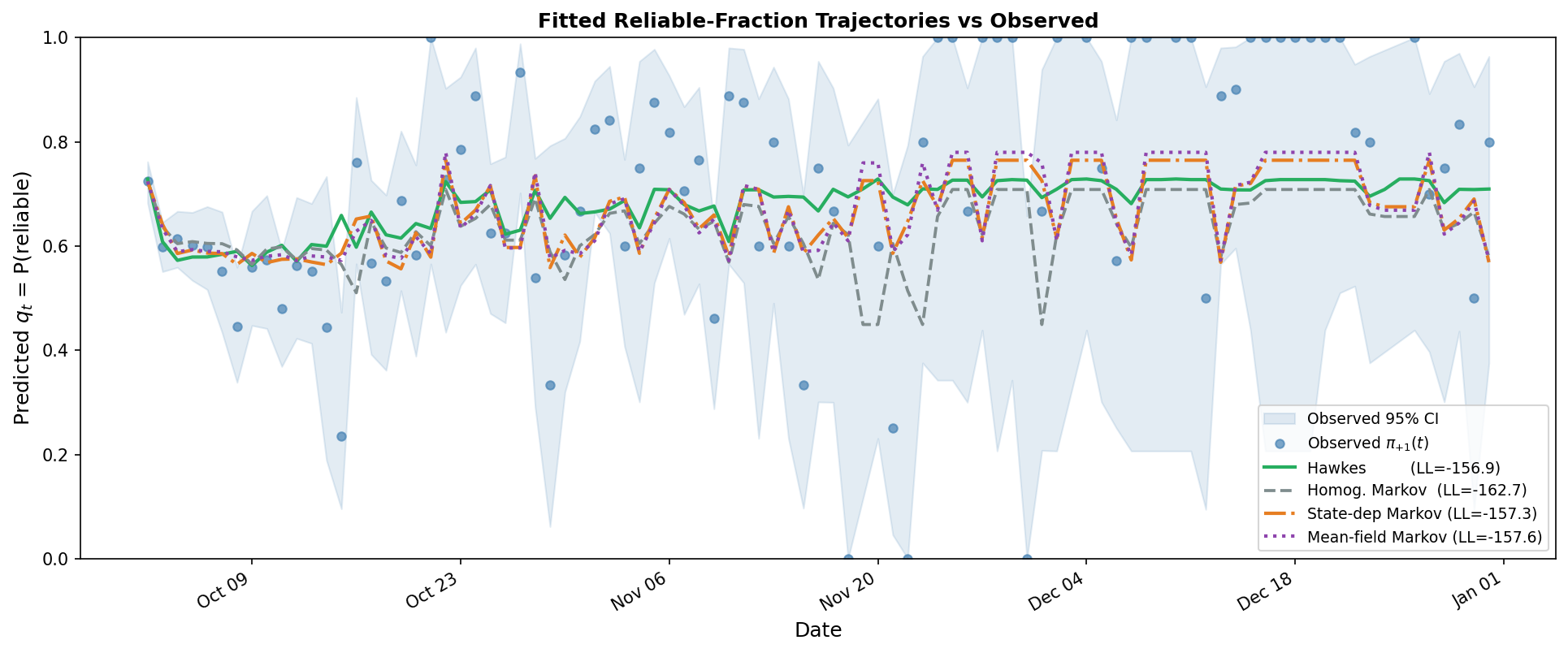}
\caption{Fitted reliable-fraction $q_t$ for all four Part-B candidates on
Las Vegas 2017, against the observed $\pipos(t)$ with Wilson 95\% bands.}
\label{fig:partB-fitted}
\end{figure}

\paragraph{Fitted-vs-observed proportions and AIC summary.}
All four candidate models track the empirical $\pipos$ similarly after
day~5; the difference is concentrated in the first week (the shock window),
which is exactly where the regime-switching matrix earns its extra
parameters. Figure~\ref{fig:partB-fitted} shows the four trajectories on
one panel.

\subsection{Cross-event replication: Hurricane Harvey 2017}
\label{sec:results-harvey}

We re-fit the same Part-A and Part-B pipelines on the Hurricane Harvey 2017
event (CC-News + NewsGuard, same loader configuration, single keyword TOML
swap). Headline numbers are summarized in
Tables~\ref{tab:harvey-partA} and~\ref{tab:harvey-partB}.

\begin{table}[H]
\centering
\caption{Part A on Hurricane Harvey 2017. Pure Hawkes-AR1 wins; the hybrid
provides no improvement because the hybrid optimizer collapses the IHP
component (the fitted $A$ is essentially zero), reflecting that Harvey's
coverage builds endogenously over multiple days rather than through a
sharp day-0 shock.}
\label{tab:harvey-partA}
\begin{tabular}{lrrrrr}
\toprule
Model & \# Params & Log-Lik. & AIC & BIC \\
\midrule
Standard Poisson         & $1$ & $-5343.00$ & $10688.00$ & $10691.02$ \\
Inhomogeneous Poisson    & $3$ & $-4539.47$ & $9084.94$  & $9094.00$  \\
\textbf{Hawkes (AR-1 limit)} & $\mathbf{2}$ & $\mathbf{-905.20}$  & $\mathbf{1814.41}$  & $\mathbf{1820.44}$ \\
Hybrid IHP+Hawkes (AR-1) & $4$ & $-904.56$  & $1817.13$  & $1829.20$  \\
\bottomrule
\end{tabular}
\end{table}

\begin{table}[H]
\centering
\caption{Part B on Hurricane Harvey 2017. State-dependent Markov is again the
AIC winner, but the homogeneous baseline is much closer than on Las Vegas:
the system spends little time in the rare regime, so regime-specific
corrections buy less.}
\label{tab:harvey-partB}
\begin{tabular}{lrrrr}
\toprule
Model & \# Params & Log-Lik. & AIC & BIC \\
\midrule
Multivariate Hawkes        & $6$ & $-126.39$ & $264.79$ & $278.69$ \\
Homogeneous Markov         & $2$ & $-124.27$ & $252.54$ & $257.17$ \\
\textbf{State-dependent Markov} & $\mathbf{4}$ & $\mathbf{-121.53}$ & $\mathbf{251.06}$ & $\mathbf{260.33}$ \\
Mean-field Markov          & $4$ & $-122.79$ & $253.59$ & $262.85$ \\
\bottomrule
\end{tabular}
\end{table}

The Harvey replication is qualitative confirmation that the methodology is
generic: same loader, same fitting code, same scoring rule, same diagnostics.
The specific winner of Part~A is event-dependent as Las Vegas's day-0 spike
demands the hybrid's exogenous-trigger term, while Harvey's gradual
endogenous build-up is fit well by pure Hawkes alone. The Part-B winner
remains the state-dependent Markov, although once again the models are nearly 
indistinguishable, but the gap to the homogeneous baseline
shrinks because Harvey spends fewer days in the neg-majority regime.

\paragraph{A keyword-filter caveat for Harvey.}
\sloppy
The two events are not on equal keyword footing. The Las Vegas filter
(\texttt{events/\allowbreak las\_vegas\_\allowbreak shooting\_2017.toml}) keys on event-specific
phrases such as ``mandalay bay'', ``stephen paddock'', ``route 91 harvest'',
and ``deadliest mass shooting'', which admit only articles substantively
about the event. The Harvey filter
(\texttt{events/\allowbreak hurricane\_\allowbreak harvey\_2017.toml}) is necessarily broader
(``harvey damage'', ``harvey recovery'', ``harvey relief'', ``harvey
texas'', ``houston flooding'', ``texas flooding'') and therefore admits
articles that mention the storm only incidentally, particularly
financial/insurance/energy reporting (Q3 disruptions, sector exposures,
reinsurance impact). This noise plausibly contributes to several Harvey
results: (i) the absence of a sharp day-0 spike, since incidental mentions
arrive throughout the window rather than being concentrated on the day of
the event, which in turn collapses the hybrid's exogenous-trigger
component (the fitted IHP amplitude on Harvey is essentially zero) and
leaves pure Hawkes-AR1 as the AIC winner; (ii) a flatter, longer count
tail that is fit comfortably by a Hawkes branching ratio of $\hat n\approx
0.95$ (very close to critical); and (iii) a shrunken regime-switching
benefit on the reliability split, since incidental business reporting
likely comes from a different reliability mix than substantive event
coverage and dilutes the signal that drives the rare neg-majority regime
on Las Vegas. Thus, the cross-event comparison is more of amethodological 
replication, and not a head-to-head comparison between the two events.

\section{Discussion}
\label{sec:discussion}

\paragraph{Headline takeaways.}
Two structural conclusions survive the comparison.
\begin{enumerate}
  \item For the count process, a hybrid IHP\,+\,Hawkes is the right model on
        events with a sharp exogenous shock. The hybrid decomposition
        matches the underlying mechanism: an exogenous trigger handles the
        first-day burst, and an endogenous self-excitation kernel carries
        the subsequent decay. On Las Vegas the two timescales are
        cleanly identifiable
        (IHP half-life $\approx 1.75$ days, Hawkes kernel half-life
        $\approx 8.2$ days).
  \item For the reliability split, a state-dependent (regime-switching)
        Markov chain is the most parameter-efficient model and gives the
        cleanest physical reading: a \emph{mean-reverting reliability
        ecology} around $\pipos\approx 0.63$. When the system slips into the
        rare neg-majority regime (only the first few shock days), the
        next-day transition probabilities push aggressively back toward
        reliable dominance.
\end{enumerate}

\paragraph{Hawkes and the Markov variants are essentially tied on fit; AIC selects by parameter count.}
On the binomial-conditional scale, all three non-null models are within
$\approx 0.6$ nats of each other on raw log-likelihood and sit
$\approx 5\text{--}6$ nats above the homogeneous Markov null
(\S\ref{sec:results-partB-LV}). The AIC verdict that places
state-dependent and mean-field Markov ahead of multivariate Hawkes is
therefore driven not by a meaningful gap in fit quality but by parameter
accounting: Hawkes pays a $+4$ AIC penalty for its $6$ parameters vs.\
the Markov variants' $4$, and that penalty exceeds the
$\approx 0.4$-nat LL gain Hawkes provides. Two structural reasons explain
why Hawkes does not earn its extra parameters back on this objective:
(i) Hawkes spends parameter capacity on predicting the \emph{total} count,
which is conditioned out by \eqref{eq:cond-binomial};
(ii) $\approx\!86\%$ of Las Vegas days are pos-majority
($74$ of $86$ non-empty days), so a single homogeneous $P$ already
captures the dominant regime, and the regime-specific tweak in 2b adds
value \emph{exactly} on the days where it is needed: the initial
neg-majority shock window. 

\paragraph{Two model choices answer two different questions.}
The Markov family parameterizes the conditional distribution we score on,
and so wins the AIC competition for forecasting tomorrow's reliability mix.
The multivariate Hawkes parameterizes cross-excitation between the two
streams, and so wins on \emph{interpretation} questions of the form ``what
kind of article triggers what kind of article tomorrow?''. The branching
matrix entries $n_{\mathrm{pos},\mathrm{neg}}=0.32$ and
$n_{\mathrm{neg},\mathrm{pos}}=0.16$ are not visible from the Markov fit at
all. Choice of model should follow the question.

\paragraph{Strengths.}
The contribution we are most confident in is the methodological one: a
shared conditional binomial likelihood that is honest about what each model
predicts and puts Hawkes and Markov on a single AIC scale. The hybrid
IHP\,+\,Hawkes count model and the mean-field Markov split model both nest
their respective null benchmarks cleanly, so the LRTs in
Tables~\ref{tab:partA-LRT} and~\ref{tab:partB-LRT} are exact.

\paragraph{Limitations.}
\begin{itemize}
  \item The binary $\pm 1$ reliability label loses the magnitude of the
        underlying NewsGuard score; a $92$ and a $55$ are both ``reliable''.
  \item The data labels are imbalanced. There are significantly more ``reliable" than 
        ``unreliable" labels; this could be the true underlying cause of the mean-reverting
        behavior, rather than the dynamic we've discussed.
  \item Day-level binning hides intraday bursts; a 3\,pm press conference is
        invisible.
  \item Single-event window for the headline analysis. The Hurricane Harvey
        replication is partial and not explicitly comparable as ``single-shock" news story. 
  \item Weekly seasonality is real (lag-7 residual autocorrelation
        $\rho_7=0.61$ on Las Vegas) but not folded into the headline model.
        $\Delta\mathrm{AIC}\approx -148$ left on the table, per
        \S\ref{sec:results-partA-LV}.
  \item The Hawkes synthetic-recovery sanity check passes at $T=500$ with
        $\approx\!20\%$ max relative error, which is acceptable for our
        purposes but not tight; real-data Hawkes parameter estimates carry
        that uncertainty.
  \item Keyword-filter precision varies across events. The Las Vegas
        filter is tightly event-specific (``mandalay bay'', ``stephen
        paddock'', ``route 91 harvest''); the Harvey filter is necessarily
        broader (``harvey damage'', ``harvey relief'', ``houston
        flooding'') and admits articles that reference the storm only
        incidentally, including financial/insurance/energy coverage. This
        confounds the cross-event comparison in \S\ref{sec:results-harvey}
        where some of the structural difference attributed to the
        \emph{event} (gradual build-up vs.\ sharp day-0 shock) is
        plausibly driven by the \emph{filter} (broad vs.\ narrow). A
        defensible cross-event verdict would require a keyword pipeline
        of matched precision, e.g.\ via a downstream classifier that
        rejects incidental mentions.
\end{itemize}

\paragraph{Verification, validation, and extensions.}
Beyond the synthetic recovery already reported, the natural validation
exercises are: (a) parameter recovery on a wider grid of synthetic
parameters; (b) posterior-predictive simulation of $N(t)$ and $\pipos(t)$
trajectories; (c) held-out-window forecasting (fit on days $0$--$60$,
forecast days $61$--$90$). Promising model extensions include:

\begin{itemize}
  \item \emph{McKean--Vlasov / distribution-dependent SDE.} Replace the
        discrete-time Markov chain on $\pipos(t)$ with a continuous-time SDE
        whose drift depends on the empirical measure of $\pipos$. This
        generalizes the mean-field Markov to a true distribution-dependent
        process and would let us derive analytic expressions for first-passage
        times, equilibrium distributions, and rare-event probabilities.
  \item \emph{Continuous reliability scores.} Drop the median threshold and
        model the raw NewsGuard score as a real-valued mark on each event,
        recovering within-class heterogeneity that the binary label hides.
  \item \emph{Cross-event replication beyond Harvey.} Running the same fitting
        pipeline on other notable events within the CC News dataset 
        (Nice 2016, Syrian civil war, Zika 2016, etc.), could expose further patterns/be used
        as a potential classification tool for species of article lifetimes.
  \item \emph{Joint count + split + weekly model.} Combine the multivariate
        Hawkes count layer with the regime-switching Markov split layer and
        add day-of-week multipliers (which the Las Vegas Part-A diagnostic
        suggests are worth $\Delta\mathrm{AIC}\approx -148$). The likelihood
        for such a joint model factorizes into a count piece and a binomial
        piece, both of which we already know how to fit.
\end{itemize}

\paragraph{Summary.}
A news cascade can be modelled as two coupled stochastic
processes on the same event timeline. The hybrid IHP\,+\,Hawkes wins on the
count process precisely because it matches the two-mechanism structure of a
breaking-news cycle: an exogenous shock followed by endogenous amplification.
The Hawkes profile-likelihood plateau identifies an AR(1) limit, which is a
first-order Markov chain, and so a Markov chain on the reliability state is
the structurally correct discrete analogue for Part B. Within that family,
the regime-switching variant wins on AIC and reveals a mean-reverting dynamic
towards a stable reliable-majority equilibrium.



\begin{thebibliography}{9}

\bibitem{CraneSornette2008}
R.~Crane and D.~Sornette,
\textit{Robust dynamic classes revealed by measuring the response function of a social system},
Proceedings of the National Academy of Sciences, 105(41):15649--15653, 2008.

\bibitem{Zhao2015SEISMIC}
Q.~Zhao, M.~A.~Erdogdu, H.~Y.~He, A.~Rajaraman, and J.~Leskovec,
\textit{SEISMIC: A self-exciting point process model for predicting tweet popularity},
in Proceedings of KDD '15, 1513--1522, 2015.

\bibitem{MishraRizoiuXie2016}
S.~Mishra, M.-A.~Rizoiu, and L.~Xie,
\textit{Feature driven and point process approaches for popularity prediction},
in Proceedings of CIKM '16, 1069--1078, 2016.

\bibitem{DaleyVereJones2003}
D.~J.~Daley and D.~Vere-Jones,
\textit{An Introduction to the Theory of Point Processes, Volume I: Elementary Theory and Methods},
2nd ed., Springer, 2003.

\bibitem{VosoughiRoyAral2018}
S.~Vosoughi, D.~Roy, and S.~Aral,
\textit{The spread of true and false news online},
Science, 359(6380):1146--1151, 2018.

\bibitem{AllcottGentzkow2017}
H.~Allcott and M.~Gentzkow,
\textit{Social media and fake news in the 2016 election},
Journal of Economic Perspectives, 31(2):211--236, 2017.

\bibitem{Tambuscio2015}
M.~Tambuscio, G.~Ruffo, A.~Flammini, and F.~Menczer,
\textit{Fact-checking effect on viral hoaxes: A model of misinformation spread in social networks},
in Proceedings of the 24th International Conference on World Wide Web Companion, 977--982, 2015.

\end{thebibliography}
\end{document}